\documentclass[reprint,
superscriptaddress,
longbibliography,
amsmath,
amssymb,
aps
]{revtex4-2}

\usepackage{epsfig,amsmath,amssymb,bm,epsf,psfrag,bbm}
\usepackage{graphicx}
\usepackage{dcolumn}
\usepackage{bm}
\usepackage{amsbsy}
\usepackage{amsthm}
\usepackage{color, comment, verbatim}
\usepackage{ragged2e}
\usepackage{subfigure}
\usepackage{enumerate}
\usepackage{chngcntr}
\usepackage[T1]{fontenc}
\usepackage[english]{babel}
\usepackage[utf8]{inputenc}

\usepackage{dsfont} 

\usepackage{tikz}
\usepackage[toc,page]{appendix}

\usepackage[colorlinks,breaklinks]{hyperref} 
\usepackage{xcolor}

\usepackage{titlesec}

\makeatletter
\newcommand\org@hypertarget{}
\let\org@hypertarget\hypertarget
\renewcommand\hypertarget[2]{%
  \Hy@raisedlink{\org@hypertarget{#1}{}}#2%
  }
\makeatother

\hypersetup{
	bookmarksnumbered,
	pdfstartview={FitH},
	citecolor={darkgreen},
	linkcolor={darkred},
	urlcolor={darkblue},
	pdfpagemode={UseOutlines}}
\definecolor{darkgreen}{RGB}{50,190,50}
\definecolor{darkblue}{RGB}{0,0,190}
\definecolor{darkred}{RGB}{238,0,0}

\usepackage{booktabs}
\usepackage{tablefootnote}
\usepackage{threeparttable}
\usepackage{color,soul}

\usepackage{natbib} 
\setcitestyle{sectionbib}
\usepackage{bibunits}
\defaultbibliographystyle{apsrev4-1fixed_with_article_titles_full_names}
\defaultbibliography{reference_list.bib}

\title{
Phase conjugation with spatially incoherent light in complex media}
\makeatletter\let\Title\@title\makeatother

\begin{document}

\title{\Title}

\author{YoonSeok Baek}
 \email{yoonseok.baek@lkb.ens.fr}
    \affiliation{Laboratoire Kastler Brossel, ENS–Universite PSL, CNRS, Sorbonne Université, Collège de France, 24 Rue Lhomond, F-75005 Paris, France}
    
\author{Hilton B. de Aguiar}
    \affiliation{Laboratoire Kastler Brossel, ENS–Universite PSL, CNRS, Sorbonne Université, Collège de France, 24 Rue Lhomond, F-75005 Paris, France}
    
\author{Sylvain Gigan}
    \affiliation{Laboratoire Kastler Brossel, ENS–Universite PSL, CNRS, Sorbonne Université, Collège de France, 24 Rue Lhomond, F-75005 Paris, France}
    
\begin{bibunit}
\begin{abstract}
Shaping light deep inside complex media, such as biological tissue, is critical to many research fields. 
Although the coherent control of scattered light via wavefront shaping has made significant advances in addressing this challenge, controlling light over extended or multiple targets without physical access inside a medium remains elusive.
Here we present a phase conjugation method for spatially incoherent light, 
which enables the non-invasive light control based on incoherent emission from multiple target positions.
Our method characterizes the scattering responses of hidden sources by retrieving mutually incoherent scattered fields from speckle patterns.
By time-reversing scattered fluorescence with digital phase conjugation, we experimentally demonstrate focusing of light on individual and multiple targets.
We also demonstrate maximum energy delivery to an extended target through a scattering medium by exploiting transmission eigenchannels.
This paves the way to control light propagation in complex media using incoherent contrasts mechanisms.
\end{abstract}

\maketitle

\section{Main}
Delivering optical energy and transmitting information through complex media remains an important challenge in many fields of studies,
including optical manipulation \cite{vcivzmar2010}, deep-tissue imaging \cite{ntziachristos2010, kubby2019} and optogenetics \cite{boyden2005, yoon2015}. 
In recent years, it has been shown that the coherent control of scattered light can manipulate spatial, spectral and temporal distributions of light in scattering media \cite{vellekoop2007, mosk2012, horstmeyer2015, rotter2017, cao2022}.
However, such capabilities are greatly limited without physical access inside a medium because the scattering response to target position is difficult to characterize.
As a result, non-invasive light control over extended or multiple targets remains elusive despite being crucial for real-world applications.
Optimizing incident wavefront based on a feedback signal \cite{katz2014, daniel2019, boniface2019, boniface2020, li2020, rauer2022, thompson2016, tian2022} is mostly limited to focusing on a single isolated target, and even then it has limitations that require numerous changes of the wavefront. 
While time-reversal or phase conjugation techniques \cite{yaqoob2008, cui2010, hsieh2010, xu2011, vellekoop2012, judkewitz2013, ma2014, zhou2014, ruan2015, ruan2017, yang2019, aizik2022} allow for effective light delivery to an optical or virtual source,
they cannot individually control light on multiple targets.

Here we address these challenges by extending phase conjugation to spatially incoherent light.
Our approach utilizes incoherent emission from multiple targets.
We first characterize scattering responses of these hidden sources by retrieving mutually incoherent fields from spatially modulated speckle patterns.
The retrieved fields are related to the field transmission matrix \cite{popoff2010}, and their phase conjugation enables light control over the desired positions.
We demonstrate this experimentally by focusing light on individual and multiple fluorescent targets through a scattering medium.
Finally, we show that transmission eigenchannels can be identified by decomposing the incoherent fields and demonstrate maximum energy delivery to a hidden extended target.

\section{Results}
\begin{figure}[ht]
	\centering
	\includegraphics[width=1.0\columnwidth]{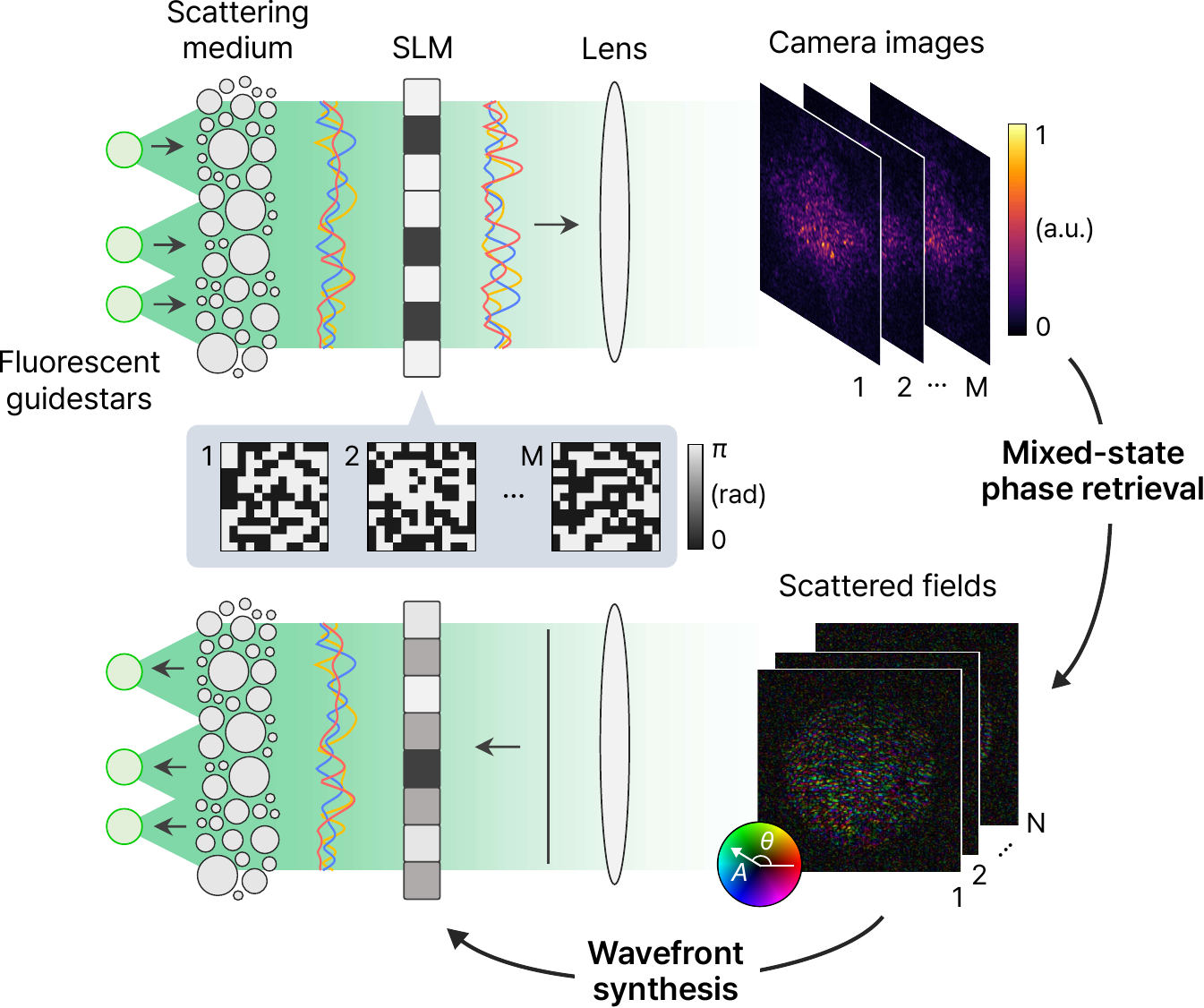}
	\caption{\textbf{Schematic of phase conjugation with incoherent fluorescence}: Multiple fluorescent guidestars are hidden behind a scattering medium. The scattered fluorescence, composed of mutually incoherent speckle fields, is modulated by an SLM. The modulated fluorescence is Fourier transformed by a lens and results in the incoherently added speckle patterns on the camera. Then mixed-state phase retrieval recovers a set of scattered fields, whose phase conjugation enables targeted light control.}
	\label{fig.schematic}
\end{figure}

A schematic of the experiment is illustrated in Fig.~\ref{fig.schematic}. 
We consider a scenario where multiple fluorescent targets are hidden by a scattering medium and act as guidestars.
Fluorescence emitted by these guidestars is scattered, resulting in an incoherent addition of speckle patterns on the camera.
Under this condition, we aim to deliver light back to each of the guidestars by time-reversing the scattered fluorescence.
Our approach consists of retrieving multiple incoherent fields that compose the fluorescence, and using them to generate phase-conjugated beams.
To this end, we introduce wavefront modulation of the scattered fluorescence with a spatial light modulator (SLM). 
This modulation induces changes in the measured incoherent speckle patterns, providing information to retrieve the scattered fields that will later be used for phase conjugation.
To explain the retrieval process in detail, we introduce a partial field transmission matrix $\bm{T}$ whose input and output are fields at the SLM plane and guidestar positions, respectively, $\bm{E}^{guidestar}=\bm{T}\bm{E}^{SLM}$. 
According to the time-reversal symmetry, the scattering of fluorescence emitted by $N$ guidestars at the same wavelength can be expressed by the rows of the transmission matrix, $\bm{T}=[\bm{t}_1,\dots,\bm{t}_N]^\top$, where $\bm{t}_n$ represents the scattered field that corresponds to an individual guidestar. 
For the sake of simplicity, we will assume that the brightness of each guidestar is the same (see Supplementary Information Section {\ref*{SI.brightness}} for general cases with different brightness).
Then, the incoherent speckle pattern measured by the camera is expressed as $\bm{I}_0 = \textnormal{diag}{\left[(\bm{TF})^{\dagger}\bm{TF}\right]}$, where $\bm{F}$ represents the discrete Fourier transform between the SLM and camera plane (see Supplementary Information Section \ref*{SI.matrix} for the derivation).
If we apply $M$ different wavefront modulations using the SLM, and express the $m$th modulation as a diagonal matrix $\bm{S}_{m}$, the incoherent speckle pattern after the modulation is 
\begin{equation}
    \bm{I}_m = \textnormal{diag}{\left[\left(\bm{TS}_{m}\bm{F}\right)^\dagger\bm{TS}_{m}\bm{F}\right]}.
    \label{eq.image}
\end{equation}

To find a set of scattered fields that satisfy Eq.~{\ref{eq.image}}, we used an iterative approach that minimizes the error between the measured and predicted $\bm{I}_m$. We began by making an initial guess of $\bm{t}_n$ and applying the wavefront modulation numerically according to Eq.~{\ref{eq.image}}. Then, following the mixed-state reconstruction {\cite{thibault2013}}, we corrected the amplitude parts of the modulated speckle fields at the camera plane, using the ratio of the measured to the predicted $\bm{I}_m$.
Next, we compensated for the wavefront modulation and updated the guess of $\bm{t}_n$. These steps were repeated for all the wavefront modulations, resulting in a maximum likelihood estimation of $\bm{T}$
(see Supplementary Information Section {\ref*{SI.PR}} for more information). 
We note that there is inherent ambiguity in determining $\bm{T}$, as $\bm{T}$ and its unitary transformation are indistinguishable based on intensity.
This can be confirmed by replacing $\bm{T}$ with $\bm{UT}$ in Eq.~\ref{eq.image}.
For this reason, the scattered fields are retrieved as a mixture of $\bm{t}_n$:
\begin{equation}
    \bm{H}=\bm{UT},
    \label{eq.u}
\end{equation}
where $\bm{H}$ is a set of retrieved fields $\bm{h}_n$, $\bm{H}=[\bm{h}_1, ... , \bm{h}_N]^\top$, and $\bm{U}$ is an arbitrary unitary matrix. 
We further note that Eq.{~\ref{eq.u}} arises from the nature of mutual incoherence, rather than from the reconstruction method. Despite this ambiguity, the retrieved fields offer unique capabilities for phase conjugation, as we show below.

\subsection*{Incoherent phase conjugation}

The time-reversal of the entire scattered fluorescence will regenerate light at hidden sources, creating foci on the entire targets.
One way to accomplish this is through the ensemble average of the phase-conjugated scattered fields $\bm{t}_n^*$. 
Alternatively, we chose to use the ensemble average of $\bm{h}_n^*$ because it gives the identical phase conjugation result (see Supplementary Information Section {\ref*{SI.incoherentOPC}}).
In both cases, the ensemble average results in the incoherent sum of the phase conjugation of the scattered fields, which can be realized by shaping either coherent or incoherent light. We will refer to this approach as ``Incoherent phase conjugation'' for simplicity, regardless of the coherence of light used to generate the phase-conjugated fields (see Discussion for more information).

\begin{figure*}[ht]
	\centering
	\includegraphics[width=1.9\columnwidth]{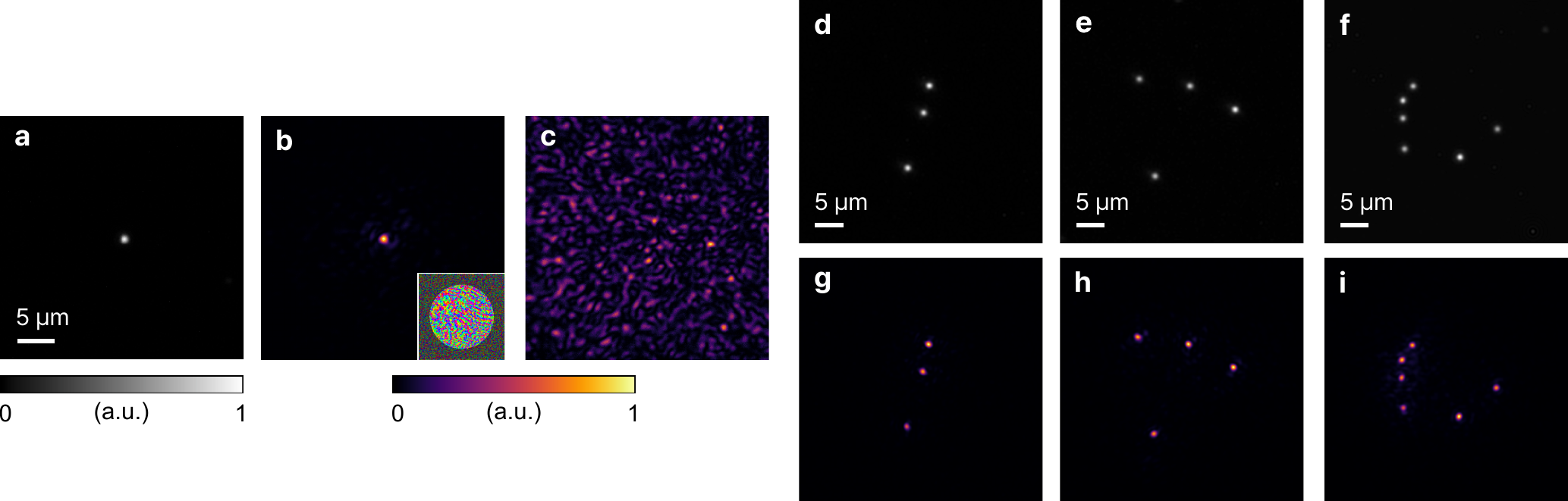}
	\caption{\textbf{Incoherent phase conjugation for multiple targets}: \textbf{a}, The fluorescence image of an 1 $\mu$m bead taken from the side without a scattering medium. \textbf{b}, Intensity at the target plane generated by phase conjugation based on the scattered fluorescence field. (Inset) The phase of the scattered field shown in the HSV colormap. The central highlighted part is used for the phase conjugation. \textbf{c}, Intensity at the target plane generated by a random wavefront. \textbf{d--f}, The fluorescence images of multiple beads hidden behind the scattering medium. \textbf{g--i}, Intensity at the target plane with incoherent phase conjugation of scattered fields.}
	\label{fig.average}
\end{figure*}

To demonstrate this incoherent phase conjugation, we introduced several 1 $\mu$m fluorescent beads as guidestars behind the scattering medium.
We retrieved the multiple scattered fields according to the number of the guidestars.
In our experiments, each 1 $\mu$m bead was considered as an individual guidestar as the speckle grain size at the target plane was $\sim$0.9 $\mu$m. 
To implement the incoherent phase conjugation, we used the SLM and a collimated laser beam to generate $N$ different phase-conjugated fields of $\bm{h}_n^*$ over time and measured their time-averaged response for phase (see Methods and Discussion).
To evaluate the performance of the phase conjugation, we first conducted an experiment with a single fluorescent bead (Fig.~\ref{fig.average}a). 
By phase-conjugating the scattered field, we observed a strong focus on the bead (Fig.~\ref{fig.average}b).
This is in clear contrast with the random speckle generated by a beam with a random wavefront (Fig.~\ref{fig.average}c).
The enhancement factor, defined as the ratio between the optimized focus intensity and mean background intensity, was $\sim$4,400.
We then placed multiple fluorescent beads (Fig.~\ref{fig.average}d--f) behind the scattering medium.
By incoherently phase conjugating the scattered fields, we successfully generated foci at every guidestar positions (Fig.~\ref{fig.average}g--i). Despite the minimal spectral memory effect \cite{vesga2019}, we were also able to excite the bead through the scattering medium by shaping the excitation beam using the retrieved field at the emission wavelength(Fig.~\ref{fig.SI_475}).

\subsection*{Selective focusing on individual targets}
In order to selectively focus on individual targets, it is necessary to demix the individual fields $\bm{t}_n$ from their mixture $\bm{h}_n$.
We note that $\bm{t}_{n}$ is not strictly orthogonal, and thus the orthogonalization of $\bm{h}_n$ can not be a solution. 
Our solution was to directly invert Eq.~\ref{eq.u} by finding $\bm{U}$.
To this end, we utilized the memory effect \cite{freund1988, feng1988}, where neighboring guidestars generate correlated speckle patterns.
Specifically, we iteratively applied a random unitary transformation to the retrieved fields $\bm{H}$, such that the correlation between the transformed speckle patterns is maximized (see Supplementary Information Section \ref*{SI.demix}). 

\begin{figure*}[ht]
	\centering
	\includegraphics[width=1.9\columnwidth]{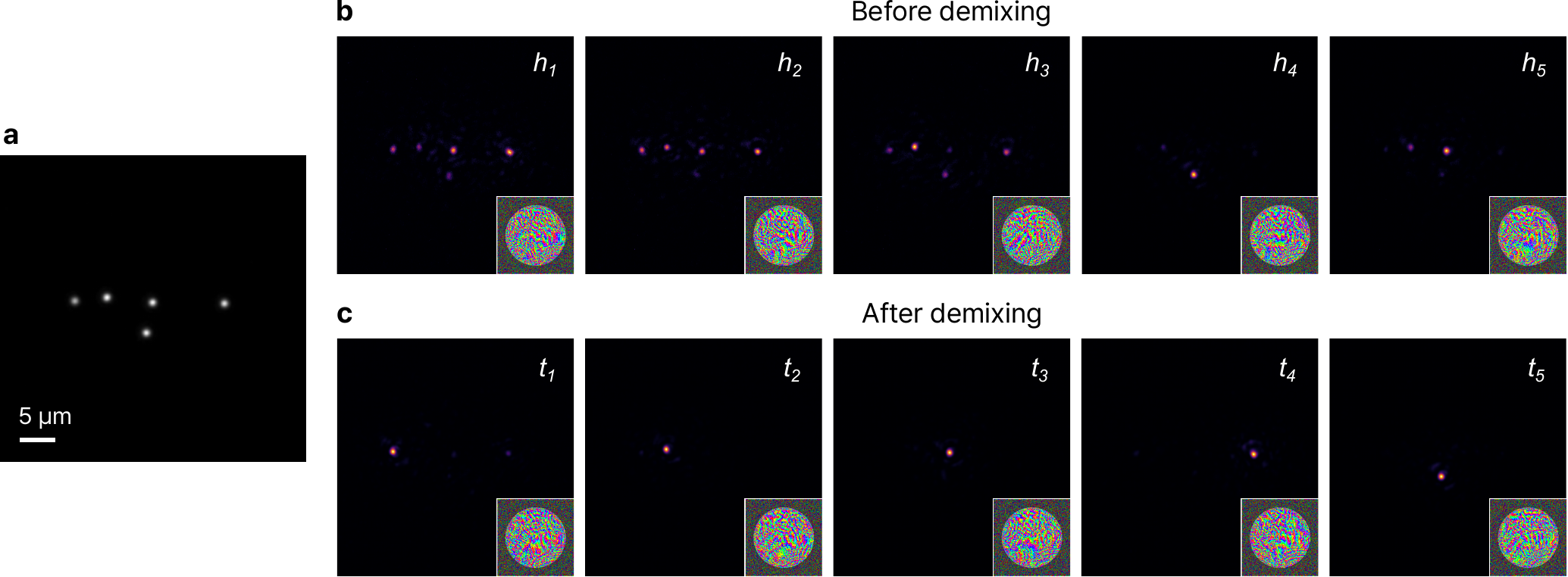}
	\caption{\textbf{Selective focusing on individual targets}: \textbf{a}, The fluorescence image of a target comprised of 1 $\mu$m beads. \textbf{b},\textbf{c}, Intensity at the target plane after the phase conjugation of individual scattered fields before (\textbf{b}) and after (\textbf{c}) the demixing process. (Insets) The phase of the scattered fields used for phase conjugation.}
	\label{fig.selective}
\end{figure*}

Figure~\ref{fig.selective} shows the experimental result with 5 fluorescent beads. When the scattered fields $\bm{h}_n$ are directly used for phase conjugation, each phase conjugation generated foci on several guidestars with different intensities (Fig.~\ref{fig.selective}b).
The demixed fields, on the other hand, generated a focus on a single guidestar, showing that $\bm{t}_n$ is successfully recovered (Fig.~\ref{fig.selective}c). 
The memory effect range in this experiment, defined as the full width at half maximum of speckle cross-correlation, was 5 $\mu$m, which is much smaller than the spatial extent of the guidestars.
This result shows that the selective focusing via demixing is possible as long as a pair of guidestars lies within the memory effect range.

\subsection*{Targeted energy delivery}
Maximum energy delivery through scattering media requires an eigenchannel of $\bm{T}$ \cite{vellekoop2008, kim2012}. 
The transmission eigenchannels correspond to the singular vectors of $\bm{T}$, and the first singular vector with the largest singular value, delivers the maximum energy to the target.
Although the direct access to $\bm{T}$ is not always possible, the transmission eigenchannels of $\bm{T}$ can be found using $\bm{H}$.
This is because the eigenchannels of $\bm{T}$ and $\bm{H}$ are identical because $\bm{H}^\dagger\bm{H}=\bm{T}^\dagger\bm{T}$ according to Eq.~\ref{eq.u}. 
Thus, we can deliver the maximum energy to extended targets using the first singular vector of $\bm{H}$.

\begin{figure*}[ht]
	\centering
	\includegraphics[width=1.8\columnwidth]{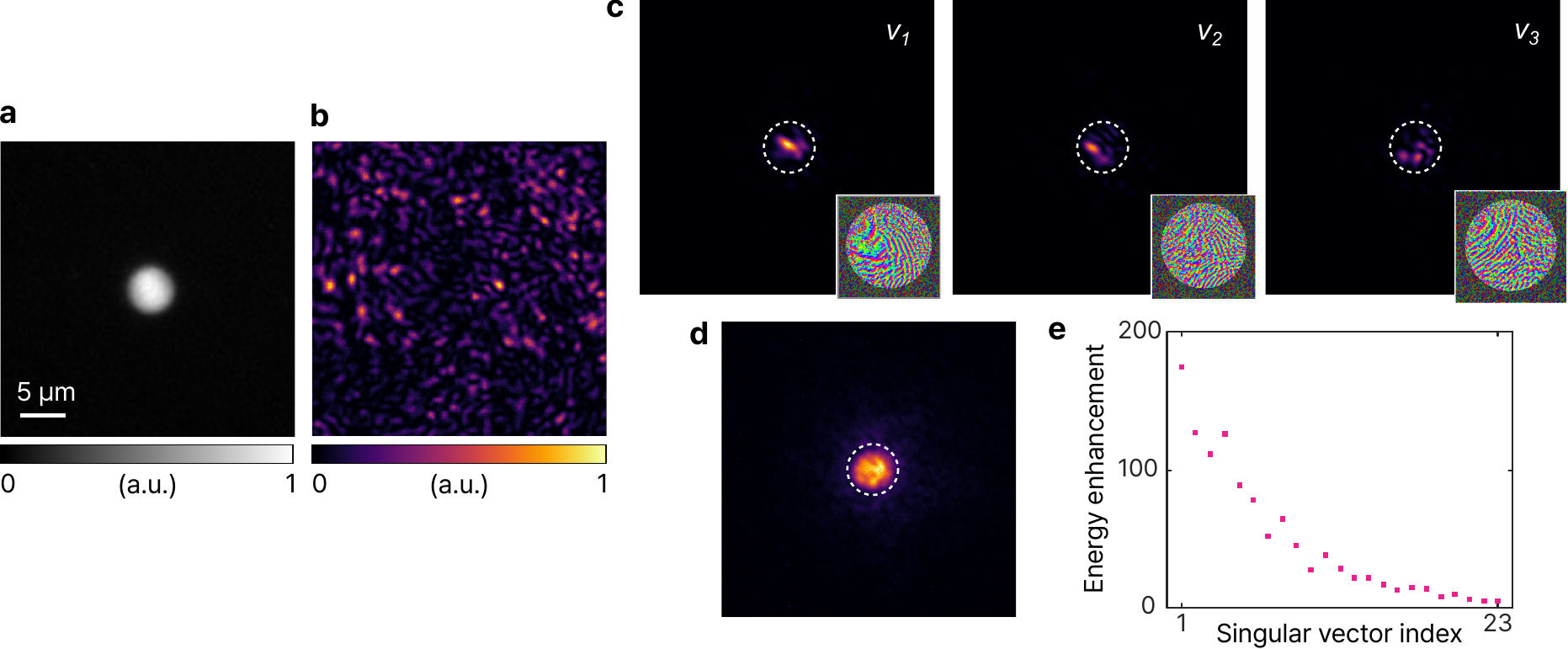}
	\caption{\textbf{Targeted energy delivery}: \textbf{a}, The fluorescence image of an extended target hidden behind a scattering medium. The image was taken from the side without the scattering medium. \textbf{b}, Intensity at the target plane when a random phase pattern is displayed on the SLM. \textbf{c}, Intensity at the target plane using the first 3 singular vectors of $\bm{H}$. The images are normalized for the result of the first singular vector. \textbf{d}, The sum of all the results using 23 singular vectors. \textbf{e}, Enhancement of the energy delivered to the target compared to random realizations. Dash circles in (\textbf{c}) and (\textbf{d}) indicate the boundary of the target.}
	\label{fig.svd}
\end{figure*}

To demonstrate the targeted energy delivery, we placed a 5 $\mu$m fluorescent ink droplet behind the scattering medium (Fig.~\ref{fig.svd}a).
Based on the size of the target, we estimated the number of incoherent fields and retrieved 23 scattered fields.
We then performed the singular value decomposition of the retrieved fields $\bm{H}$.
Finally, we injected fields that corresponds to the singular vectors $\bm{v}_n$ and observed the energy delivered to the target.
When a random wavefront is injected to the scattering medium, a speckle pattern is generated at the target plane (Fig.~\ref{fig.svd}b).
In contrast, the singular vectors produce intensity distributions highly concentrated on the target (Fig.~\ref{fig.svd}c). 
By summing the results of all the singular vectors, we confirmed that the energy is delivered only to the target area (Fig.~\ref{fig.svd}d).
The first singular vector $\bm{v}_1$ shows a 174-times increase in the energy on the target, compared to random wavefronts. 
The enhancement decays with the singular vector index [Fig.~\ref{fig.svd}(e)].
We observed that the values are not perfectly sorted in a descending order, which we believe is due to the numerical error in the retrieved fields and to the use of the phase-only SLM.

\section{Discussion and conclusion}
We have presented an approach to control light in scattering media without the physical access to a target plane by extending phase conjugation to spatially incoherent light.
We have demonstrated focusing and maximum energy delivery to extended fluorescent targets through a scattering medium.
Our method, like other digital phase conjugation or wavefront shaping techniques, can provide reliable focusing results with a high contrast provided a sufficiently large number of incident modes are controlled {\cite{mosk2012}}. Nonetheless, what distinguishes our approach is its ability to tackle mutually incoherent fields of scattered fluorescence.
Another important aspect is that it does not require the precise alignment between the camera and SLM \cite{hillman2013}, since the scattered fields are retrieved at the SLM plane.
Its principle of characterizing the scattering response is entirely passive, as it does not alter the emission of guidestars, as opposed to to techniques that modulates the excitation wavefront (e.g. \cite{boniface2020}).
It only requires that the emission from the sources be constant on average over the measurement of $\bm{I}_m$ and thus unaffected by the setup geometry.

To further demonstrate the capability of the proposed method under realistic conditions, we conducted an additional experiment. Specifically, we employed a volumetric scattering medium composed of parafilm layers, and we increased the number of guidestars. More information on the experiment can be found in Supplementary Information Section {\ref*{SI.parafilm}}. The results showed that the proposed method was able to produce high-contrast foci on 13 and 22 target positions (Fig.{~\ref{fig.SI_parafilm}}), demonstrating the potential of our method for practical applications.

For the proposed method, it is important to estimate the number of mutually incoherent waves, $N$. 
This is because the underestimation of $N$ results in imperfect reconstruction of the scattered fields. 
We note that the overestimation is allowed because it results in redundant reconstruction (see Fig.~\ref{fig.SI_number}).
Nevertheless, it is recommended to use the smallest possible value of $N$ for the minimal measurements and computation time.
There are different methods to estimate $N$. 
The contrast of fluorescence speckle is an useful indicator for $N$, as it decreases as $\sqrt{N}$ \cite{goodman2007}.
It is also possible to find $N$ by analyzing the error in the mixed-state phase retrieval or the singular value distribution \cite{popoff2011} for different values of $N$.
In our demonstrations, we did not consider the spectral degrees of freedom because narrow spectral responses were measured by using interference filters.
If the detection bandwidth is greater than the spectral memory effect range, different spectral components should be considered in estimating $N$.
This will allow the proposed technique to handle situations where hidden sources emit at different wavelengths.

Another important consideration is the number of modulation $M$ required for the field retrieval, which scales linearly with $N$. 
In experiments, reliable phase conjugation results were obtained when $M\geq6N$ (see Supplementary Information Section \ref*{SI.PR}).
This linearity can be attributed to the multiplexed information in the intensity of multiple incoherent fields.
We emphasize that $M$ scales with the number of incoherent waves $N$ not with the number of controlled input modes of the SLM.
Recent advances in phase retrieval \cite{candes2015wirtinger, candes2015matrix} show that few measurements are sufficient in retrieving a coherent field ($N = 1$).
In this regard, we believe that in principle even fewer $M$ may be used for our method.

In our proof-of-principle experiments, we used the simple algorithms for the retrieval and demixing of the scattered fields.
The performance of the algorithms can be enhanced by incorporating constraints, convex optimization \cite{candes2015matrix} or the generalized memory effect \cite{osnabrugge2017}.
The total measurement time can be reduced by designing a setup with minimal energy loss, and by using a sensitive detector, such as an EMCCD.

The current implementation of incoherent phase conjugation is based on the alternation of multiple coherent fields. This method can accurately generate the incoherent response of the time-reversed fluorescence, although it requires some integration time \mbox{\cite{de1986}}. Importantly, the required integration time can be reduced to few microseconds by employing fast spatial light modulators \mbox{\cite{rodenburg2014, tzang2019}}. 
Alternatively, the incoherent phase conjugation can be implemented by directly shaping an incoherent source \mbox{\cite{barre2022}}. In this case, the incoherent response will be obtained at a timescale greater than the coherence time of the source.

In conclusion, our method enables versatile light control over extended or multiple targets using incoherent contrast mechanisms.
The concept can be applied to different incoherent emissions, such as spontaneous Raman scattering \cite{thompson2016,tian2022}, and a wide range of photoluminescence \cite{kim2019}.
Furthermore, it enables the passive characterization of a transmission matrix, opening up the possibility to generalized light control using transmission-matrix-based operators \cite{cao2022}.
We envision that the proposed approach will enable targeted light delivery through thick biological tissue, facilitating biomedical applications, such as optogenetic stimulation and phototherapy.

\section{Methods}

\subsection*{Experimental setup}
The experimental setup is shown in Fig.~\ref{fig.SI_setup}.
A laser diode ($\lambda$ = 488 nm, LP488-SF20G, Thorlabs) was used to excite fluorescence. 
The excitation beam was delivered to guidestars by a lens (L1, f = 100 mm) and an objective lens (Plan N 20$\times$ 0.4, Olympus).
To monitor the guidestars and phase conjugation, a dichroic mirror (DMLP490R, Thorlabs), a lens (L2, f = 200 mm), a bandpass filter (FL532-10, Thorlabs), and a camera (acA5472-17um, Basler) were placed on the side without a scattering medium.
The guidestars were fluorescent ink mixed with UV glue (NOA 68, Norland), and fluorescent beads (F8823, Invitrogen) immersed in glycerol.
A scattering medium was a 220-grit ground glass diffuser, placed approximately 170 $\mu$m away from the guidestars.
On the detection side, the scattered fluorescence was collected by an objective lens (MPlan N 50$\times$ 0.75, Olympus) and two lenses (L3, f = 75 mm; L4, f = 150 mm). 
An SLM (X10468-04, Hamamtsu) and a linear polarizer were used to modulate the fluorescence.
The modulated fluorescence is Fourier transformed by lenses (L5, f = 100 mm; L6, f = 200 mm; L7, f = 250 mm) and then captured by an sCMOS camera (PCO.edge 5.5, PCO) with bandpass filters (BP2, FL532-3 and FBH520-40, Thorlabs). For the fluorescent beads we used, these filters transmit 5.35\% of the total fluorescence signal. 
An iris was placed between L6 and L7 to adjust the speckle grain size at the camera.
For phase conjugation, a laser ($\lambda$ = 532 nm, Compass 215M-50, Coherent) was collimated using a 5 $\mu$m pinhole and a lens (L8, f = 6 mm).
The collimated laser was then shaped by the SLM to generate a phase-conjugated beam, which propagated back to the medium.
A flip mirror was used to switch between the fluorescence detection and phase conjugation.

\subsection*{Phase-conjugated beam generation}
A phase-conjugated beam is generated using the collimated laser beam and the SLM.
The collimated beam is shaped to the phase conjugate of a given scattered field $\bm{E}_{scattered}$ by displaying a phase pattern that corresponds to $-\textnormal{arg}\left(\bm{E}_{scattered}\right)$ on the SLM.
The resultant phase-conjugated beam propagates back through the scattering medium, retracing the scattering paths of fluorescence.

\section*{Acknowledgement}
This research was supported by H2020 Future and Emerging Technologies (863203), European Research Council (724473), and Basic Science Research Program through the National Research Foundation of Korea (NRF) funded by the Ministry of Education (2022R1A6A3A03072108).


\newpage

\putbib
\end{bibunit}

\hspace{5cm}
\newpage


\newpage
\onecolumngrid
\newpage
\begin{center}
{\large \textbf{Supplementary Information: \\[2mm] \Title}}\\[1cm]

\end{center}

\renewcommand\thesection{\arabic{section}}
\setcounter{section}{0}
\setcounter{equation}{0}
\setcounter{figure}{0}
\setcounter{table}{0}
\setcounter{page}{1}

\renewcommand{\theequation}{S\arabic{equation}}
\renewcommand{\thefigure}{S\arabic{figure}}
\renewcommand{\thetable}{S\arabic{table}}

\titleformat{\section}
    [block]{\normalfont\bfseries\normalsize}{\rlap{\thesection}}{0em}
    {\hspace{0.035\linewidth}}

\begin{bibunit}

\section{Matrix representation of {\textit{I}}{\textsubscript{0}} and {\textit{I}}{\textsubscript{m}}}
\label{SI.matrix}
The intensity speckle measured at the camera plane can be written as
\begin{equation}
    I_0(\bm{k})
    =\sum_{n=1}^N{\left|\mathcal{F}[t_n\left(\bm{r}\right)]\right|^2}(\bm{k})
    =\sum_{n=1}^N{\left|\tilde{t}_n\left(\bm{k}\right)\right|^2},
\label{eq.I0}
\end{equation}
where $t_n(\bm{r})$ and $\tilde{t}_n(\bm{k})$ is the $n$th scattered field at the SLM plane and camera plane, respectively, and $\mathcal{F}$ is the Fourier transform. To express Eq.~{\ref{eq.I0}} in a matrix form, we utilize $\bm{T}$ to express $t_n(\bm{r})$ and $\tilde{t}_n(\bm{k})$. First, we can express $t_n(\bm{r})$ as the rows of $\bm{T}$, such that $\bm{T}_{ij}=t_i(r_j)$. Similarly, $\tilde{t}_n(\bm{k})$ can be expressed as the rows of $\bm{TF}$, where $\bm{F}$ is a Fourier transform matrix, such that $(\bm{TF})_{ij}=\tilde{t}_i(k_j)$. If we consider a matrix $(\bm{TF})^\dagger\bm{TF}$, its elements can be expressed as $[(\bm{TF})^\dagger\bm{TF}]_{ij}=\sum_{n=1}^N{\tilde{t}^*_n(k_i)\tilde{t}_n(k_j)}$. Therefore, its main diagonal is given by $[(\bm{TF})^\dagger\bm{TF}]_{ii}=\sum_{n=1}^N{\left|\tilde{t}_n(k_i)\right|^2}$, which is equivalent to $I_0(k_i)$. Finally, we can rewrite Eq.~{\ref{eq.I0}} as $\bm{I}_0 = \textnormal{diag}{\left[(\bm{TF})^{\dagger}\bm{TF}\right]}$ by expressing the main diagonal of a matrix using the function $\textnormal{diag}$. Equation~{\ref{eq.image}} can be obtained by following a similar process as described above, but by replacing $\bm{T}$ with $\bm{T}\bm{S}_{m}.$

\section{Mixed-state phase retrieval} \label{SI.PR}
\begin{figure*}[hp]
	\centering
	\includegraphics[width=0.35\columnwidth]{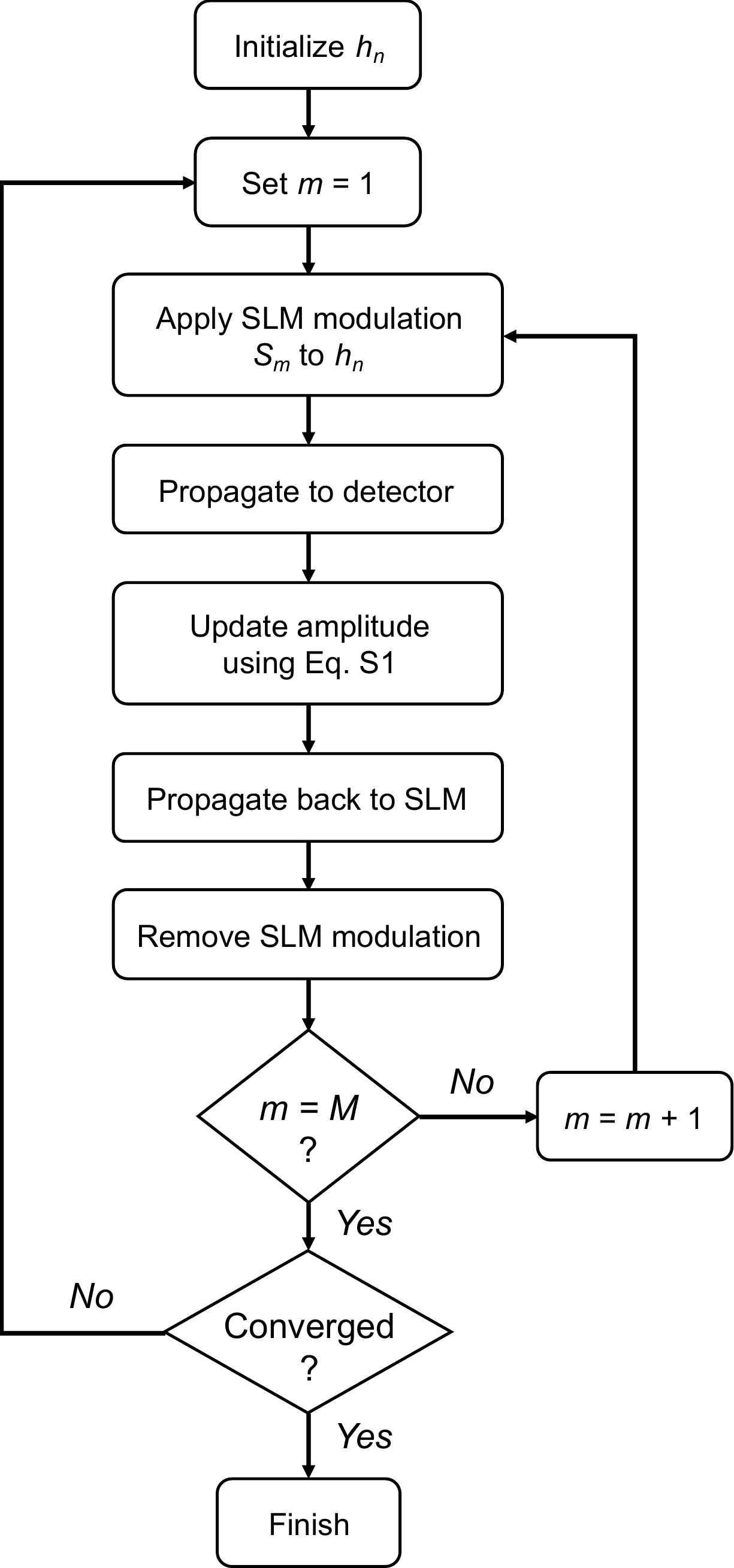}
	\caption{\textbf{Flowchart of the mixed-state phase retrieval}}
	\label{fig.SI_flowchart}
\end{figure*}

\begin{figure*}[hp]
	\centering
	\includegraphics[width=0.35\columnwidth]{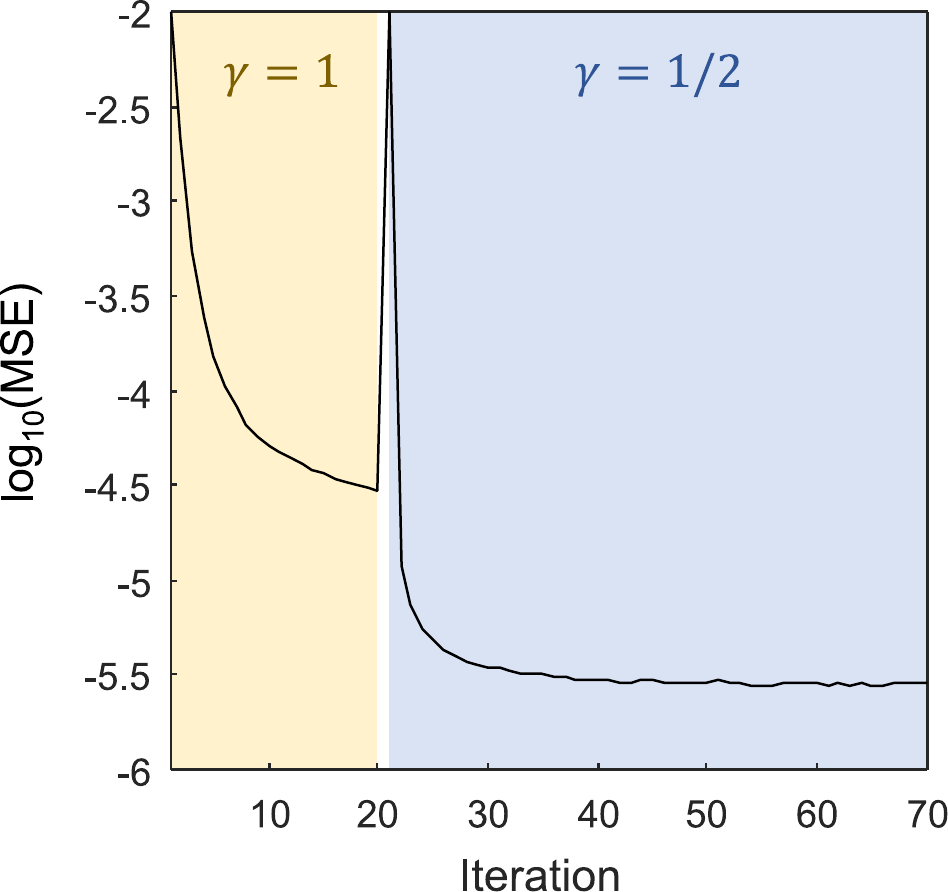}
	\caption{\textbf{Experimental mean squared error}}
	\label{fig.SI_MSE}
\end{figure*}

Our phase-retrieval algorithm is designed to find $N$ mutually incoherent fields at the SLM plane from $M$ modulated intensity images. 
We initialize the algorithm by letting the scattered fields $\bm{h}_n$ as $N$ complex Gaussian random fields. 
Then we apply the wavefront modulation $\bm{S}_m$ to the scattered fields $\bm{h}_n$.
We note that $\bm{S}_m$ is a diagonal matrix whose diagonal elements correspond to the field modulation given by the SLM.
In the experiments, SLM was divided into macro-pixels (composed of 40 $\times$ 40 pixels) having random phase values (0 or $\pi$).
After the wavefront modulation $\bm{S}_m$, the field at the camera plane is expressed as $\bm{\tilde{h}}_n^{(m)} = \left(\bm{S}_m\bm{F}\right)^{\top}\bm{h}_n$.
Then we conduct Fourier magnitude projection using an auxiliary function $\psi$:
\begin{equation}
\psi_n^{(m)}(\bm{k}) = 
\frac{\left[I_m(\bm{k})\right]^{\gamma}}{\sqrt{\sum_n{\left|\tilde{h}_n^{(m)}(\bm{k})\right|^2}}}
\tilde{h}_n^{(m)}(\bm{k}),
\label{eq.update1}
\end{equation}
where $\bm{k}$ is a coordinate in the spatial frequency domain, $\tilde{h}_n^{(m)}(\bm{k})$ is the modulated field at the camera plane, corresponding to $\bm{\tilde{h}}_n^{(m)}$, and $\gamma$ is a constant parameter. 
Then the fields are updated by compensating the wavefront modulation:
\begin{equation}
    \bm{h}_n = \left[\left(\bm{S}_m\bm{F}\right)^{\top}\right]^{-1}{\bm{\psi}}_n^{(m)},
\label{eq.update2}
\end{equation}
where $\bm{\psi}_n^{(m)}$ is the vector representation of $\psi_n^{(m)}(\bm{k})$. 
The update through Eq.~\ref{eq.update1}--\ref{eq.update2} is continued for the entire $M$ modulations. The whole process is repeated for several times to obtain consistent $\bm{h}_n$ (see Fig.{~\ref{fig.SI_flowchart}}). 
With $\gamma = 1/2$, this method can be interpreted as the maximum likelihood reconstruction and retrieves the incoherent fields \cite{thibault2013}.
However, we observed that few initial iterations with $\gamma = 1$ accelerates the convergence greatly, in both numerical simulations and experiments.
This observation agrees with the effect of the nonlinear modulus constraint that was reported in \mbox{\cite{zhang2010,zhang2007}}.
It should be noted when {$\gamma$} is greater than 1/2, the brighter area of $I_m$ carry more weight during the iteration. While it has been reported that the larger values of {$\gamma$} lead to faster convergence \mbox{\cite{zhang2010}}, we expect that the optimal value of $\gamma$ for the initial iterations would depend on the specific application. In our experiments, we used $\gamma = 1$ for the first 20 iterations and $\gamma = 1/2$ for the rest of the iterations.
We stopped the algorithm when there was no significant change in {$\bm{h}_n$}. Specifically, we evaluated the convergence by using the mean squared error (MSE) between the current and previous estimates of {$\bm{h}_n$}:
\begin{equation}
    \textnormal{MSE} = \frac{1}{NK}\sum_n^N\sum_k^K{\left|h_{nk}^{\textnormal{(current)}}-h_{nk}^{\textnormal{(previous)}}\right|}^2,
\end{equation}
where {$h_{nk}$} is the $k$th element of {$\bm{h}_n$}.
In our experiments, we stopped the algorithm when the MSE was less than $10^{-5.6}$. See Fig.{~\ref{fig.SI_MSE}} for a change of the MSE over iterations.
We confirmed numerically that the algorithm retrieves a correct set of fields for $M\geq4N$ in the absence of measurement noise.
The size of macro-pixel had almost no effect on the reconstruction, except when its size is comparable to the SLM. 
We observed that the minimum value of $M$ required for correct reconstruction increases depending on the noise level.
For the experimental results shown in the main text, we used 6--8$N$ modulations.

\section{Incoherent phase conjugation} \label{SI.incoherentOPC}
When all the mutually incoherent components of fluorescence are time-reversed, the intensity at the \textit{n}th guidestar can be expressed by multiplying the \textit{n}th row of the transmission matrix, $\bm{t}_{n}^\top$, and phase-conjugated field $\bm{t}_m^*$:
\begin{equation}
\sum_{m}\left|{\bm{t}_{n}^{\top}\bm{t}_{m}^{*}}\right|^{2}
= \bm{t}_{n}^{\top}\bm{T}^{\dagger}\bm{T}\bm{t}_{n}^*,
\label{eq.focusT}
\end{equation}
where $\ast$ denotes the complex conjugate.
For a medium with negligible reflection and absorption, $\bm{T}^{\dagger}\bm{T}\approx\mathds{1}$. Thus Eq.~\ref{eq.focusT} is simplified to $\bm{t}_{n}^{\top}\bm{t}_{n}^{*}$, which remains more or less constant regardless of \textit{n}.
As a result, the incoherent phase conjugation of $\bm{t}_n$ generates foci on the entire guidestars with roughly the same intensity.
It should be noted that $\bm{T}$ is not unitary in practice, since we are not able to measure the entire scattered light. Nevertheless, if the loss of information due to absorption or measurement is minimal, the phase conjugation can approximate the case of unitary $\bm{T}$ for complex media, producing high-contrast foci on the entire guidestars \mbox{\cite{tanter2000, horstmeyer2015}}.
Similarly, the intensity for the incoherent phase conjugation of $\bm{h}_{n}$ is expressed as,
\begin{equation}
\sum_{m}\left|{\bm{t}_{n}^{\top}\bm{h}_{m}^{*}}\right|^{2} = \bm{t}_{n}^{\top}\bm{H}^{\dagger}\bm{H}\bm{t}_{n}^*.
\label{eq.focusH}
\end{equation}
Since $\bm{H}^{\dagger}\bm{H}=\bm{T}^{\dagger}\bm{T}$ according to Eq.~\ref{eq.u}, Eq.~\ref{eq.focusH} is identical to Eq.~\ref{eq.focusT}.
Thus, the incoherent phase conjugation of $\bm{h}_n$ and $\bm{t}_n$ are identical.

\section{Additional experiment with volumetric scattering medium} \label{SI.parafilm}

In addition to the incoherent phase conjugation described in Fig.~{\ref{fig.average}}, we conducted an additional experiment with a volumetric scattering medium. In the experiment, we placed 1 $\mu$m fluorescent beads directly behind three layers of parafilm, with a total thickness of approximately 360 $\mu$m corresponding to half of the transport mean free path {\cite{boniface2019}}. 
We replaced the objective lens with a long-working-distance objective lens (MPlanFL N 100$\times$ 0.9, Olympus) and used a 10-nm bandpass filter (FLH05532-10, Thorlabs) which transmits 14.3\% of the fluorescence signal. The rest of the setup was the same. The experiment was conducted in the same manner as the experiment for Fig.~{\ref{fig.average}}. 

\begin{figure*}[hp]
        \centering
	\includegraphics[width=0.6\columnwidth]{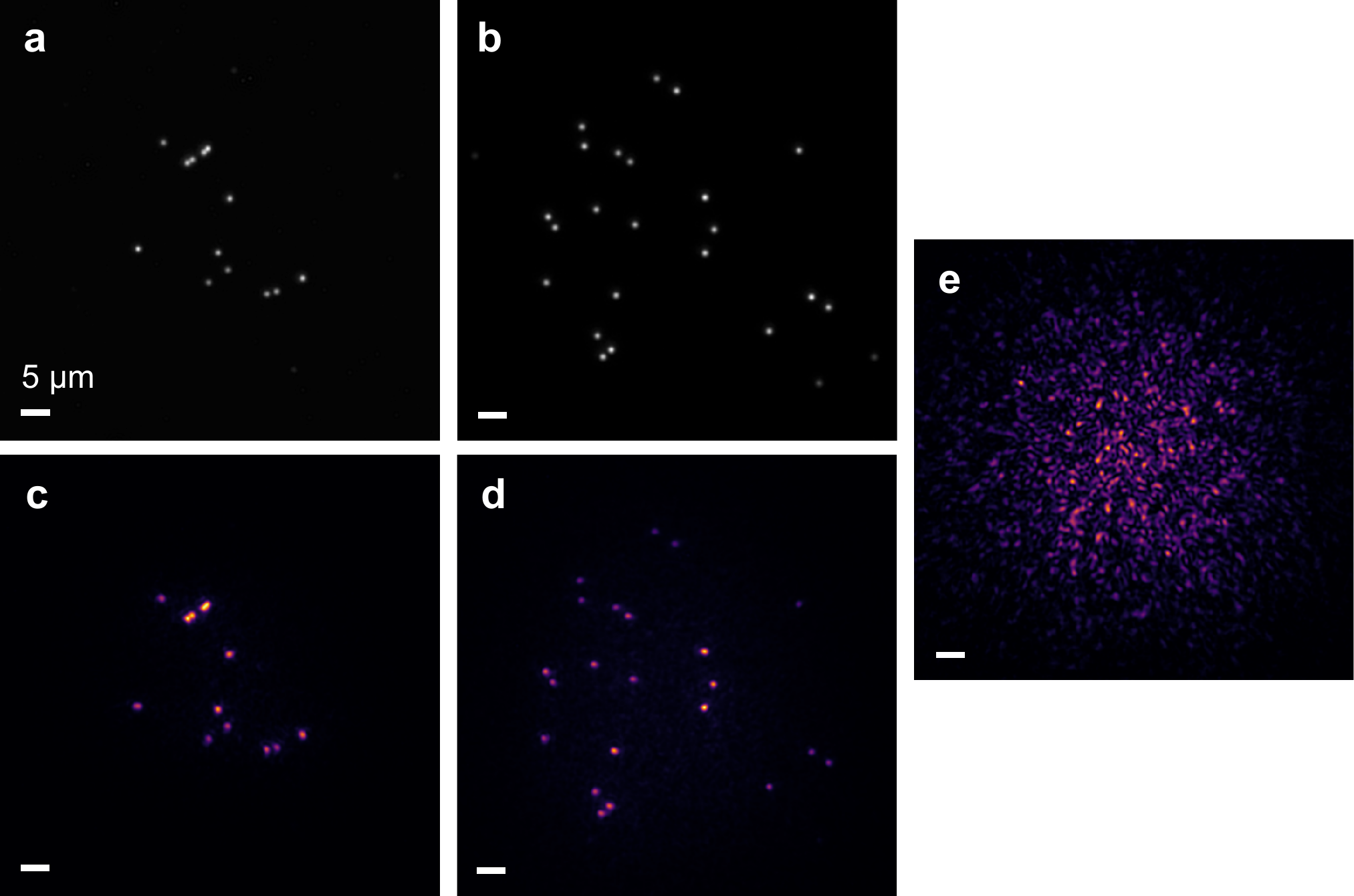}
	\caption{\textbf{Incoherent phase conjugation in volumetric scattering medium}: \textbf{a--b} Fluorescence image of beads hidden behind 3 layers of parafilm. \textbf{c--d} Results of the incoherent phase conjugation. \textbf{e} Intensity at the target plane with a random wavefront.}
	\label{fig.SI_parafilm}
\end{figure*}

The results showed that that the proposed method was able to produce high-contrast foci on 13 and 22 guidestars (Fig.~{\ref{fig.SI_parafilm}}). In Fig.~{\ref{fig.SI_parafilm}}d, we observed that foci were not generated at a few target positions located far from the center. This is due to the limited field-of-view (FoV) in which the SLM can control light. By evaluating the speckle pattern generated by a back-propagating beam with a random wavefront, we confirmed that the size of the controllable FoV was approximately 60 $\mu$m (Fig.~{\ref{fig.SI_parafilm}}e). This is smaller than the field of view of Fig.~{\ref{fig.SI_parafilm}} (75.6 $\mu$m). We note that the FoV is determined by the scattering media and the experimental setup, such as the focal length of an objective lens and the SLM pixel size.

\section{Demixing incoherent fields} \label{SI.demix}
The scattered fields $\bm{t}_n$ can be recovered by reversing Eq.~\ref{eq.u}. 
To this end, we developed an algorithm inspired by the simulated annealing \cite{kirkpatrick1983} that exploits the memory effect between the speckle patterns at the camera plane.
The algorithm starts by applying a random unitary transformation to the retrieved fields $\bm{H}$ and expressing the result at the camera plane:
\begin{equation}
    \bm{\overline{H}} = {\bm{U}_j}^{f_j}\bm{\hat{U}}\bm{H}\bm{F},
    \label{eq.inv}
\end{equation}
where $j$ is the iteration index, $\bm{U}_j$ is a random unitary matrix, $f_j$ is an exponent that decays to 0 over the iteration, and $\bm{\hat{U}}$ is the current best estimate of $\bm{U}^{-1}$. 
The decaying behavior of $f_j$ makes $\bm{U}_j$ induce smaller changes as the iteration progresses: $\lim_{j\to\infty}{\bm{U}_j}^{f_j} = \mathds{1}$. 
Then the correlation metric $C$ is calculated at every iteration,
\begin{equation}
\label{eq.metric}
    C = \sum_{n=1}^{N}\Bigl[\max\left({|\overline{h}_n|^2\star|\overline{h}_{(n\bmod N) +1}|^2}\right)\Bigr]^2,
\end{equation}
where ${\overline{h}}_n(\bm{k})$ is the unitary-transformed speckle field at the camera plane, corresponding to the $n$th row of $\bm{\overline{H}}$, and $\star$ is the cross-correlation.
If the value of $C$ is greater than the previous values,  $\bm{\hat{U}}$ is updated to ${\bm{U}_j}^{f_j}\bm{\hat{U}}$. 
If not, $\bm{\hat{U}}$ is unchanged.
At the end of the iteration, the algorithm results in $\bm{\hat{U}} \approx \bm{U}^{-1}$, and the scattered fields are recovered by inverting Eq.~\ref{eq.u}: $\bm{T}=\bm{\hat{U}}\bm{H}$.
We initialized the iteration with $\bm{\hat{U}} = \mathds{1}$ and $f_j = 100/(j+1)$, and the maximum iteration number of $10^4$.
We note that the experimental noise in $\bm{H}$ can lead to an incorrect demixing.
To mitigate this issue, we calculate $C$ only using speckles brighter than the average, $|\overline{h}_n|^2 > \mu$, where $\mu$ is the mean value of every $|\overline{h}_n|^2$. 

\section{Guidestars of varying brightness} \label{SI.brightness}
If the brightness of the guidestars is different, we can express the scattered field corresponding to $n$th guidestar as $b_n\bm{t}_n$, where $b_n$ is a constant that is the square-root of the energy emitted by the guidestar and $\bm{t}_n$ is a normalized field, such that \mbox{$\bm{t}_n^{\dagger}\bm{t}_n=1$}. Then we can rewrite the expressions in the main text by replacing $\bm{T}$ with $\bm{BT}$, where \mbox{$\bm{B}=\textnormal{diag}(b_1,...,b_n)$}. For example, the unmodulated speckle image becomes \mbox{$\bm{I}_0 = \textnormal{diag}{\left[(\bm{BTF})^{\dagger}\bm{BTF}\right]}$}. Similarly, Eq.~{\ref{eq.image}} is rewritten as
\begin{equation}
    \bm{I}_m = \textnormal{diag}{\left[\left(\bm{BTS}_{m}\bm{F}\right)^\dagger\bm{BTS}_{m}\bm{F}\right]},
    \label{eq.brightness_image}    
\end{equation}
and Eq.~{\ref{eq.u}} as
\begin{equation}
    \bm{H}=\bm{UBT}.
    \label{eq.brightness_u}
\end{equation}
We note that this generalization is related to the interpretation of the measured image and retrieved fields. Therefore, the mixed-state reconstruction and the demixing process are unaffected by the brightness of the guidestars. For incoherent phase conjugation (Supplementary Section~{\ref*{SI.incoherentOPC}}), foci will be generated on the entire guidestar with the intensities that correspond to the brightness of the guidestars. Regarding maximum energy delivery, the eigenchannel of $\bm{H}$ will preferentially couple light into the brighter guidestar positions because \mbox{$\bm{H}^\dagger\bm{H}=\bm{T}^\dagger\bm{B}^2\bm{T}$}, where \mbox{$\bm{B}^2=\textnormal{diag}(b_1^2,...,b_n^2)$}.

\section{Maximum number of guidestars} \label{SI.number}

In principle, the mixed-state algorithm can find solutions regardless of the number of guidestars $N$, provided that there are enough wavefront modulations. However, in practice, the number of guidestars could be limited because the dynamic range of a camera causes quantization of images. For the phase retrieval of coherent light, it is known that 3-bit representation of speckle can produce reliable phase retrieval results {\cite{maallo2010}}. Considering that the mixed-state reconstruction can be thought of as the information multiplexing of mutually incoherent signals, we would expect the following condition to be satisfied,
\begin{equation}
    N\times2^3\leq2^{\textnormal{(Camera bit depth)}}.
\end{equation}
For example, with a 16-bit resolution, the maximum theoretical $N$ will be 8192, although in practice it would be significantly lower. Finally, we note that even a larger number could be allowed by incorporating the quantization error into the reconstruction algorithm {\cite{yang2021}}.

\newpage

\begin{figure*}[hp]
        \centering
	\includegraphics[width=0.8\columnwidth]{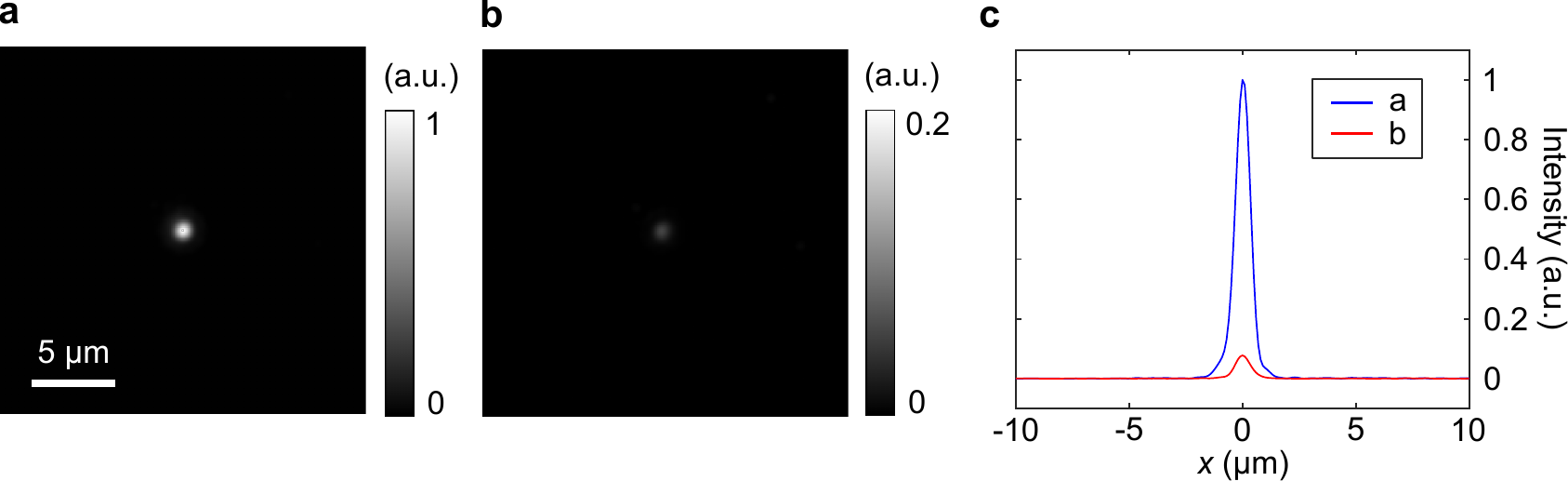}
	\caption{\textbf{Phase conjugation using an excitation beam}: \textbf{a} Fluorescence image of a bead when an excitation beam ($\lambda = 475~\mu m$) is shaped using the phase conjugation pattern of scattered fluorescence ($\lambda = 532~\mu m$). \textbf{b} Fluorescence image of the same bead when the excitation beam is shaped by a random phase pattern. \textbf{c} Intensity profiles of (a) and (b) along the horizontal line that crosses the center.}
	\label{fig.SI_475}
\end{figure*}

\begin{figure*}[hp]
	\centering
	\includegraphics[width=0.7\columnwidth]{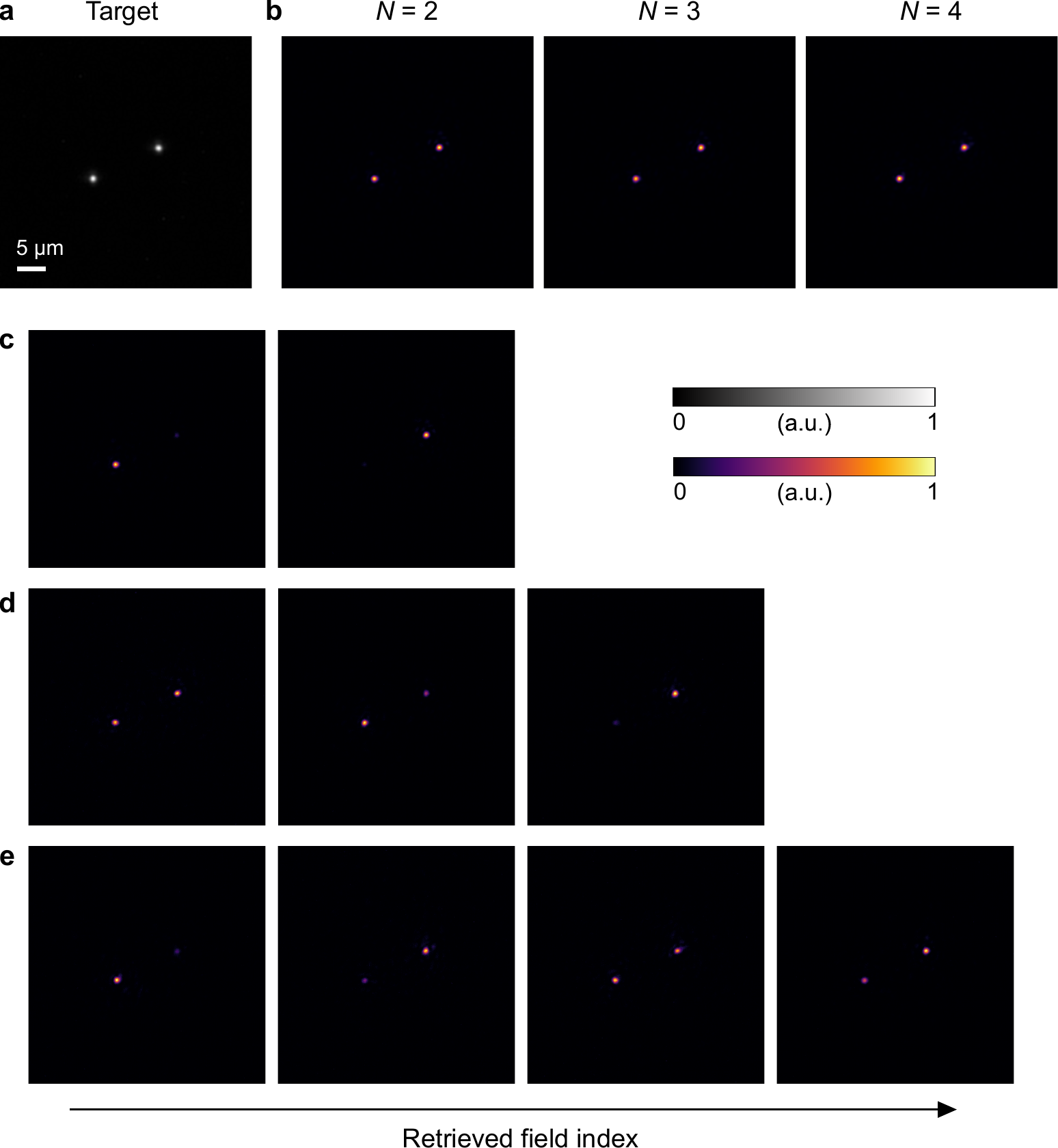}
	\caption{\textbf{Phase conjugation with different numbers of retrieved fields}: \textbf{a} Fluorescence image of hidden targets. \textbf{b} Results of incoherent phase conjugation when different numbers of scattered fields are retrieved. \textbf{c--e} Phase conjugation of individual fields when two (c), three (d), and four (e) scattered fields are retrieved.}
	\label{fig.SI_number}
\end{figure*}

\begin{figure*}[hp]
	\centering
	\includegraphics[width=0.7\columnwidth]{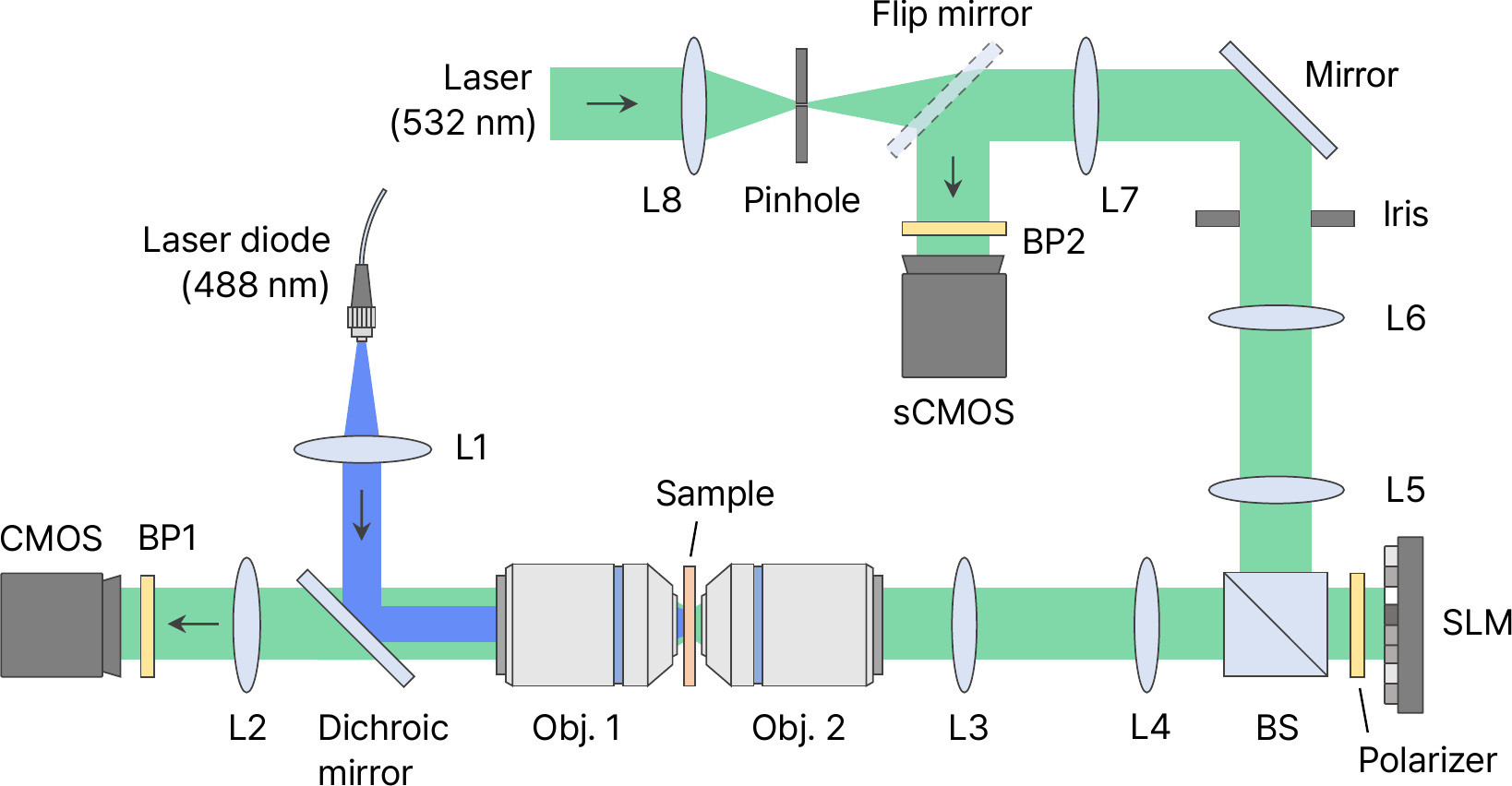}
	\caption{\textbf{Experimental setup}: L, lens; Obj., objective lens; BS, beam splitter; BP, banspass filter.}
	\label{fig.SI_setup}
\end{figure*}

\clearpage 
\putbib
\end{bibunit}

\end{document}